# Compressed $k^2$-Triples for Full-In-Memory RDF Engines

**Sandra Álvarez-García, Nieves R. Brisaboa**
Univ. of A Coruña (Spain)
salvarezg@udc.es, brisaboa@udc.es

**Javier D. Fernández, Miguel A. Martínez-Prieto**
Univ. of Valladolid (Spain), Univ. of Chile (Chile)
jfergar@infor.uva.es, migumar2@infor.uva.es

### ABSTRACT

Current "data deluge" has flooded the Web of Data with very large RDF datasets. They are hosted and queried through SPARQL endpoints which act as nodes of a semantic net built on the principles of the Linked Data project. Although this is a realistic philosophy for global data publishing, its query performance is diminished when the RDF engines (behind the endpoints) manage these huge datasets. Their indexes cannot be fully loaded in main memory, hence these systems need to perform slow disk accesses to solve SPARQL queries.

This paper addresses this problem by a compact indexed RDF structure (called **$k^2$-triples)** applying compact $k^2$-tree structures to the well-known vertical-partitioning technique. It obtains an ultra-compressed representation of large RDF graphs and allows SPARQL queries to be full-in-memory performed without decompression. We show that $k^2$-triples clearly outperforms state-of-the-art compressibility and traditional vertical-partitioning query resolution, remaining very competitive with multi-index solutions.

### Keywords

RDF compressed index, SPARQL querying, $k^2$-tree.

### INTRODUCTION

The W3C originally recommended RDF (*Resource Description Framework*) as a "foundation for processing metadata" and establishes that its broad goal is "to define a mechanism for describing resources". This conception was clearly influenced by a *document-centric* perspective of the Web. Although the current Recommendation[1] shares these original foundations, it also devises an evolution by suggesting the use of RDF "to do for machine processable information (application data) what the WWW has done for hypertext: to allow data to be processed outside the particular environment in which it was created, in a fashion that can work at Internet scale". This statement shows a perfect description of the RDF status in the current evolution of the Web, where the *Linked Data project*[2] plays a crucial role.

Linked Data is about employing RDF and HTTP to publish structured data on the Web and to connect it between different data sources. This movement has allowed the original document-centric perspective of the Web to evolve to a *data-centric* one in which the term "Web of Data" is commonly used. RDF is a cornerstone in this Web of Data providing a graph-based data model for structuring and linking data which describes facts of the world (Bizer, Heath and Berners-Lee, 2009). This knowledge is modeled by using atomic *triple* units which comprise a subject "S" (the resource being described), a predicate "P" (the property), and an object "O" (the property value). In turn, the SPARQL Recommendation[3] establishes a specific language for querying RDF data. It is based on *triple patterns, i.e.,* RDF triples in which each subject, predicate or object may be a variable. Thus, SPARQL queries contain many conjunctions of triple patterns which are resolved by graph matching.

The Web of Data comprises very large RDF datasets from diverse fields. Its size is currently estimated in 25 billion triples[4]. Thus, performance and scalability issues arise when this data needs to be managed and queried. One can think that size is not an issue because cheaper disks are available for storage purposes. However, large size not only penalizes querying but also the performance of other common processes such as RDF publication and exchange. Note that this latter one is becoming even more popular due to the increasing usage of SPARQL endpoints to perform remote SPARQL queries which transmit their results through the net. A recent work (Fernández, Martínez-Prieto and Gutiérrez, 2010) addresses this problem by understanding the logical structure of an RDF dataset. This approach, called HDT (*Header-Dictionary-Triples*), considers three components to describe an RDF dataset: 1) the *Header* contains metadata about the dataset; 2) the *Dictionary* organizes the different URIs, blanks and literals used in the dataset, and 3) the set of *Triples* encodes the pure structure of the underlying RDF Graph. HDT allows very compact RDF representations to be achieved and gives a natural division of components to facilitate RDF management in real applications. We will use these three terms in the rest of the paper to refer the corresponding component in an RDF dataset.

[1] `http://www.w3.org/TR/rdf-syntax-grammar/`

[2] `http://linkeddata.org/`

[3] `http://www.w3.org/TR/rdf-sparql-query/`

[4] `http://www4.wiwiss.fu-berlin.de/lodcloud/`

This paper deals with a compact representation of the RDF structure (*Triples*) able to be queried with SPARQL. We consider sophisticated compressed data structures that allow Triples to be efficiently loaded and queried in main memory due to its small size. Our approach, called $k^2$-triples, models the Triples component by using compressed $k^2$-trees (Brisaboa, Ladra and Navarro, 2009) originally conceived for compact representation of web graphs, but recently generalized for representing general graph databases (Álvarez, Brisaboa, Ladra and Pedreira, 2010). Thus, $k^2$-triples gives an ultra-compressed index on the structure of the RDF graph and allows SPARQL queries to be performed without prior decompression. All this process is carried out in main memory enhancing the query performance process against to state-of-the-art engines which also need to access secondary memory because of the large size of their (multiple) indexes.

The main contributions of this paper are:

- An ultra-compressed representation of RDF graphs, allowing very large datasets to fit in main memory.
- Native support for *triple pattern* queries on the compressed representation.
- Efficient performance for conjunctive queries on top of the triple patterns.

The paper is organized as follows. Next section reviews the state-of-the-art for RDF engines and describes the $k^2$-tree structure as basis to our $k^2$-triples approach. The Evaluation section compares $k^2$-triples against the state-of-the-start, focusing on compression ratios and querying times for some real-world datasets. Finally, we devise future lines of work from the current achievements.

## STATE-OF-THE-ART

RDF is a logical data model which does not limit its physical storage. This fact allows different solutions to be approached for storing RDF in an effective way. Some of them have stored RDF in a relational database and have performed SPARQL queries through SQL; C-Store (Abadi, Marcus, Madden and Hollenbach, 2007), Monet-DB[5] or Virtuoso[6] are well-known systems implementing this solution. Two basic policies are considered to transform RDF into a relational representation: (1) storing all triples in a large 3-column table `[S,P,O]`, and (2) grouping triples by predicate and defining a specific 2-column table: `[S,O]` for each one; this last technique, called *vertical-partitioning,* is based on the fact that few predicates are used to describe a dataset. A third hybrid policy combines the previous two to obtain some 3-column table clustering correlated predicates

The description of vertical-partitioning policy implicitly suggests a *subject-object* (`SO`) index for each predicate. It allows some SPARQL queries to be speeded up, but makes some others difficult: *e.g.* triple patterns with unbounded predicates. C-Store and Monet-DB, adequately tuned (Sidirourgos, Goncalves, Kersten, Nes and Manegold, 2008), provide these features on column-oriented databases. Other approaches (regarding a triple as a 3-dimensional entity) arise under this previous weakness. Hexastore (Weiss, Karras and Bernstein, 2008) and RDF-3X (Neumann and Weikum, 2010) are well-known systems which create indexes for all ordering combinations (`SPO`, `SOP`, `PSO`, `POS`, `OPS`, `OSP`). Although their main goal is achieving a global competitive performance, this index replication largely increases spatial requirements. RDF-3X reduces this effect by applying gap-compression to the leafs of the $B^+$-tree storing triples.

A recent approach, called BitMat (Atre, Chaoji, Zaki and Hendler, 2010), suggests a compressed bit-matrix structure for storing huge RDF graphs. This description shows two promising features: 1) a native compressed structure for RDF representation, and 2) the ability to manage huge graphs in contrast with the lack of scalability of some previous systems (Sidirourgos et al., 2008). Although BitMat is conceptually designed as a *bit-cube* `S×P×O`, its practical representation slices to get two-dimensional matrices: `SO` and `OS` for each predicate, `PO` for each subject and `PS` for each object. These matrices are gap-compressed by taking advantage of their sparseness. Whereas the feature 2) is demonstrated on huge datasets, the compression comparison shows that BitMat is not able to outperform RDF-3X. All these works present a common weakness already reported in experimental studies: the problem of managing large datasets and the efficiency waste of disk transfers (Sidirourgos et al., 2008).

---

[5] `http://monetdb.cwi.nl/`

[6] `http://www.openlinksw.com/dataspace/dav/wiki/Main/VOSRDF`

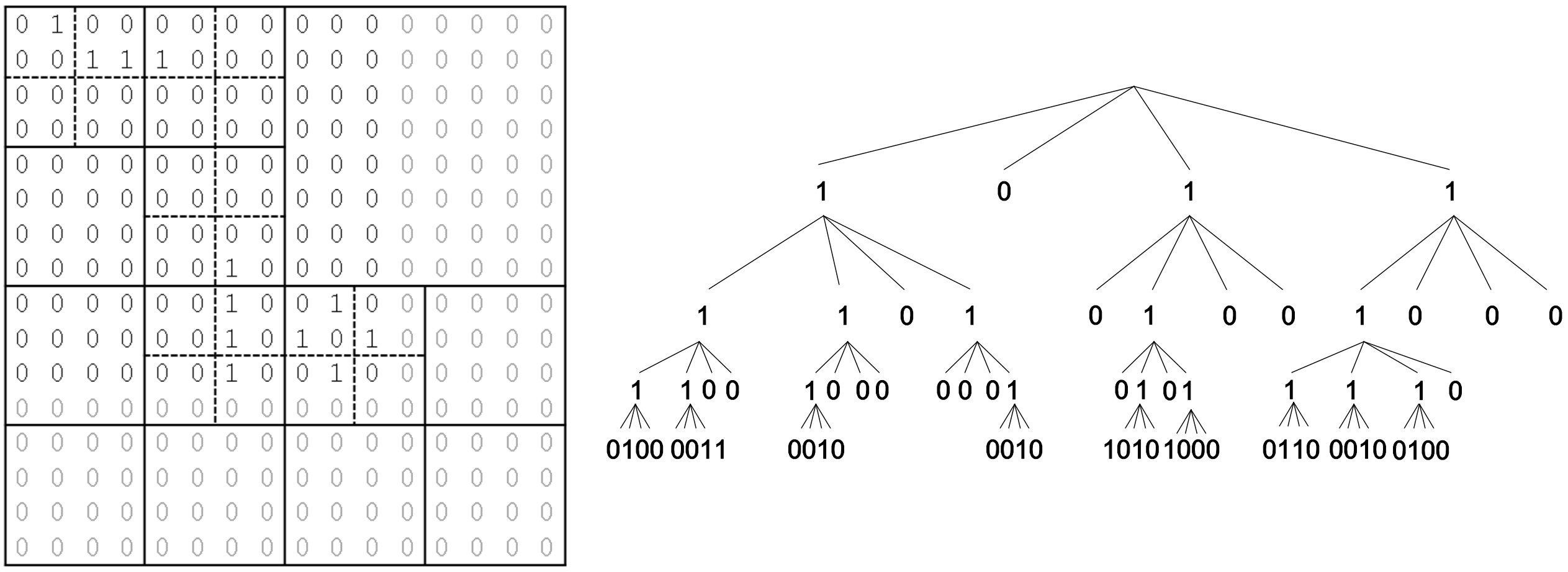

**Figure 1. Example of $k^2$-tree (k=2)**

### $k^2$-trees

$k^2$-trees (Brisaboa et al, 2009) was originally designed as a compact structure for web graph representation taking advantage of large empty areas existing in its adjacency matrix. This method achieves a very compact space and enables an efficient bidirectional navigation over the graph. It allows direct and inverse neighbors to be retrieved on the compressed structure. Additionally, $k^2$-trees also supports other navigation operations such as range queries or checking a particular cell value.

The $k^2$-tree method represents the adjacency matrix using a non-balanced $k^2$-ary tree. It subdivides the input matrix into $k^2$ submatrices of the same size. Each of those submatrices will be a child of the root node and its value will be `1` *iff* the submatrix contains at least one cell with value `1`. Otherwise, its value will be `0` and the tree decomposition ends there because this submatrix is empty. Once this first level has been built, the method proceeds recursively for each child with value `1`, until it reaches the last level of the tree where the elements corresponds to the matrix cell values. Figure 1 shows an example of an adjacency matrix of a graph, and how it is subdivided to obtain the corresponding $k^2$-tree using `k=2`.

The whole adjacency matrix is represented in a very compact way using just two bit-arrays: `T` (tree) and `L` (leaves). `T` stores all the levels except the last one, following a levelwise traversal. `L` stores the last level of the tree. An auxiliary structure is created over `T` to allow navigation through the compact representation of the tree. To find the neighbors of a node, the $k^2$-tree needs to locate which cells in a certain row or column of the adjacency matrix have a `1`. To locate those cells, we go down through the *k* children corresponding to those submatrices that overlap with the row (column) of the node of the query. This top-down traverse can be efficiently performed over the bit-arrays `T` and `L`.

All $k^2$-trees used in this work are physically built on a hybrid policy which uses values k=4, up to the level 5 of the tree, and then k=2, for the rest ones (Brisaboa et al, 2009). The leaf level is encoded using Directly Addressable Codes parameterized with b=8 (Ladra, 2011).

### $K^2$-TRIPLES

This approach provides a compact representation for the *Triples* component *w.r.t.* the philosophy of Header-Dictionary-Triples partitioning for large RDF datasets (Fernández et al.,2010). $k^2$-triples is based on the vertical-partitioning of the dataset. It allows triples to be regarded as a cloud of points distributed in the underlying two-dimensional structure of the $k^2$-tree. This way, $k^2$-triples achieves ultra-compressed representations which can be fully loaded and queried in main memory and allows us to face up the problems underlying in the state-of-the-art.

Each triple `(S,P,O)` is mapped to a set of three integer-IDs which identify their corresponding components in the Dictionary. We consider a previous experience (Atre et al., 2010) to describe the following four categories (being each one lexicographically sorted):

- Common subjects and objects (`SO`) contain all terms which play subject and object roles in the dataset. These terms are mapped to the range [`1`, |`SO`|] in order to obtain efficient cross-joins among subjects and objects.
- Subjects (S) contain all subjects which do not play an object role. These are mapped to [`|SO|+1`, `|SO|+|S|`].

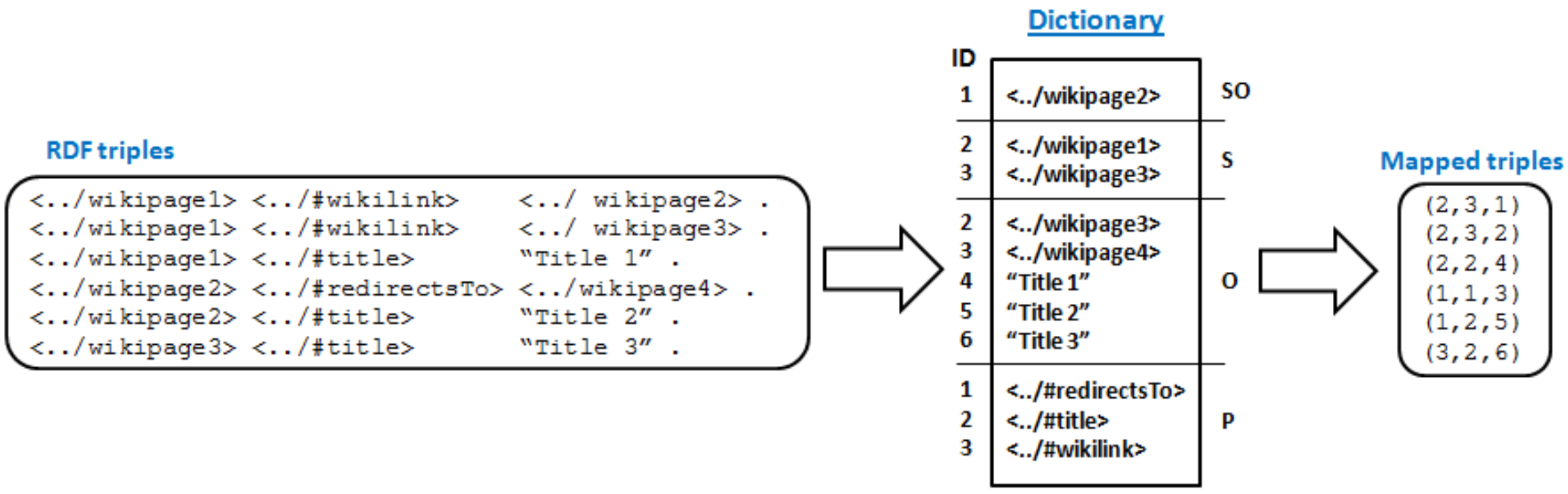

**Figure 2. Mapping from triples to IDs**

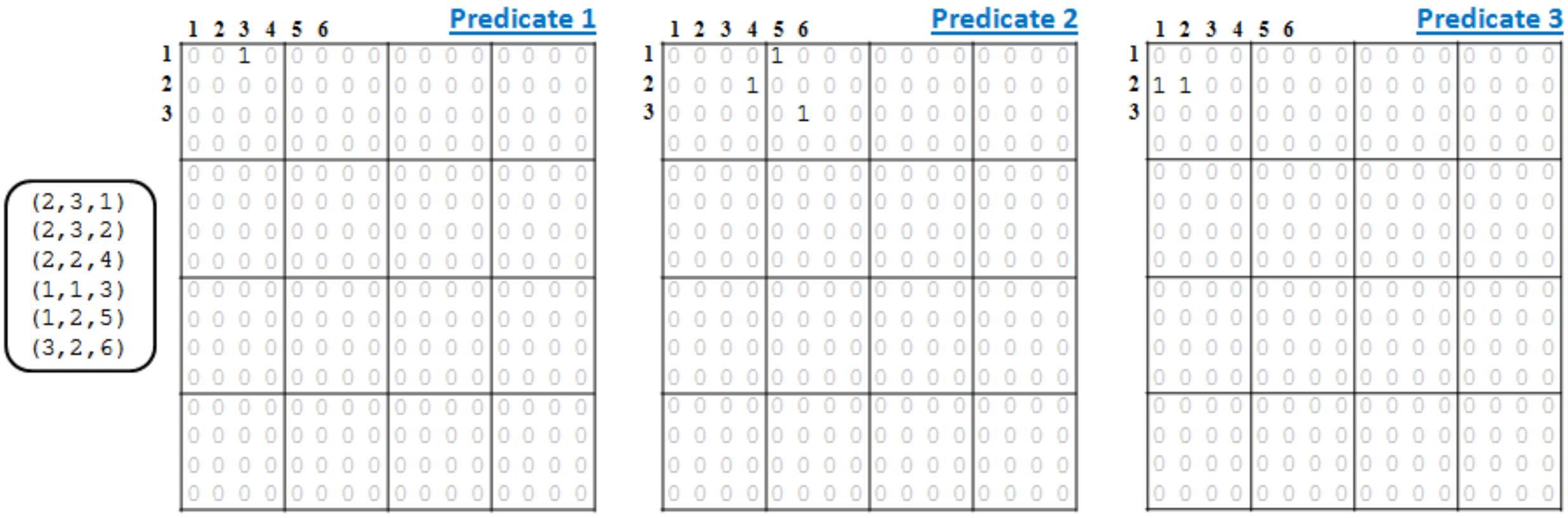

**Figure 3. Vertical-Partitioning on $k^2$-trees (k=4).**

- Objects (O) contain all objects which does not play a subject role. These are mapped to [|SO|+1, |SO|+|O|].
- Predicates (P) contain all predicates. These are mapped to [1, |P|].

Figure 2 shows an example of mapping from triples (left) to IDs (right) and the resultant Dictionary configuration for this RDF. As can be seen, `<../wikipage2>` is the unique element playing roles of subject and object and it is identified with the ID 1; subject elements are identified as 2 and 3, object elements are identified from 2 to 6, and the predicates use the IDs from 1 to 3.

### Vertical-Partitioning on $k^2$-trees

We consider the previous experiences in *vertical-partitioning* to approach *$k^2$-triples*: we use an independent $k^2$-tree for indexing all triples associated with a certain predicate. All these $k^2$-trees are designed as square binary matrices representing *subjects* as rows and *objects* as columns. Its final size is rounded up to the next power of k: $n=k^{\lceil \log_k(|SO|+\max(|S|,|O|)) \rceil}$. This extension is made by padding with 0s to the right and to the bottom. This decision does not cause a significant overhead by considering the $k^2$-tree ability to handle large areas of 0s.

Figure 3 shows how $k^2$-triples represents the dataset considered in Figure 2. Three independent $k^2$-trees are used for indexing the triples associated with each predicate. Note that only the three first rows (for the three existing subjects) and the six first columns (for the six objects) are really used in each $k^2$-tree, hence all triples are stored in these ranges. For instance, the predicate 2 takes part in three triples: (2,2,4),(1,2,5) and (3,2,6), and its corresponding $k^2$-tree stores them in the coordinates (2,4),(1,5)and (3,6), which represent the corresponding subject-object pairs.

Final $k^2$-tree configurations describe very sparse 1 distributions. This result is in line with (Fernández et al., 2010) which shows that subjects related to predicate-object and objects related to subject-predicate are very few for real-world datasets. In this scenario, $k^2$-trees arise as an effective solution due to its ability to represent these NULL areas in a compact manner.

As we explain below, all basic SPARQL queries can be answered on $k^2$-tree primitive operations. That is, all triple patterns[7] are implemented by checking points existence or by traversing column/rows:

---

[7] Triple pattern notation uses ?X values to indicate unbounded elements.

- The simple pattern (S,P,O) asks about the existence of the given triple. It is answered by checking whether the point (S,O) in the $k^2$-tree representing P contains the bit 1. In turn, the pattern (S,?P,O) is solved by performing the previous operation on all $k^2$-trees. That is, the points (S,O) are checked in all $k^2$-trees and all predicates containing 1 in this cell are returned as results.
- Patterns (S,P,?O) and (S,?P,?O) are implemented, like *direct neighbors*, as a forward navigation (from subject to object) of the $k^2$-tree. The first pattern (returning all objects related with the subject S through the predicate P) is solved by locating all 1s in the row of the subject S. The resulting list is ordered by object-ID. (S,?P,?O) is basically solved by performing the previous pattern for all predicates and retrieving all existing pairs (P,O) for the given subject S.
- Patterns (?S,P,O) and (?S,?P,O) are implemented, like *reverse neighbors*, as a backwards navigation (from object to subject) of the $k^2$-tree. The first pattern (returning all subjects related with the object O through the predicate P) is solved by locating all 1s in the column of the object O. The resulting list is ordered by subject-ID. (?S,?P,O) is basically solved by performing the previous pattern for all predicates and retrieving all existing pairs (S,P) for the given subject O.
- The pattern (?S,P,?O) retrieves all pairs (S,O) for a given predicate P. This operation returns all 1s in the chosen $k^2$-tree by performing a range search query. Finally, the unusual pattern (?S,?P,?O) retrieves all triples in the dataset by performing the previous pattern for all predicates.

The conjunction of these patterns allows more complex queries to be obtained through **join** conditions. Previous work on vertical-partitioning (Abadi et al., 2007) explains how joins on subjects can be efficiently performed because of subjects are indexed for each predicate. This also provides fast merge joins on subjects when the predicate is unbounded in the pattern. An additional index, which adds a significative overhead to the final representation size, is needed to achieve the same performance on objects. However, the main weakness of vertical-partitioning is related to queries involving unbounded predicates. It is already an expensive operation by considering that it need to access to all tables representing the given predicates, but it also requires extra I/O time to load some of these tables from secondary memory.

$k^2$-triples faces up these problems and gives native support for their solutions. On the one hand, the $k^2$-tree provides indexed access to subjects and objects allowing both subject-subject and object-object joins to be efficiently performed. Additionally, $k^2$-triples allows required merge joins to be directly performed on the sorted lists returned as result for patterns with unbounded subjects or objects. Therefore, it is also efficient for cross-joins (between subjects and objects). Note that these operations can be easily solved in the submatrix [1, |SO|]×[1, |SO|] which stores, for each predicate, all possible results for these joins. On the other hand the ultra-compressibility of $k^2$-triples allows all $k^2$-trees to fir in main memory, saving the disk transference overheads and enabling its efficient traversing to solve queries with unbounded predicates.

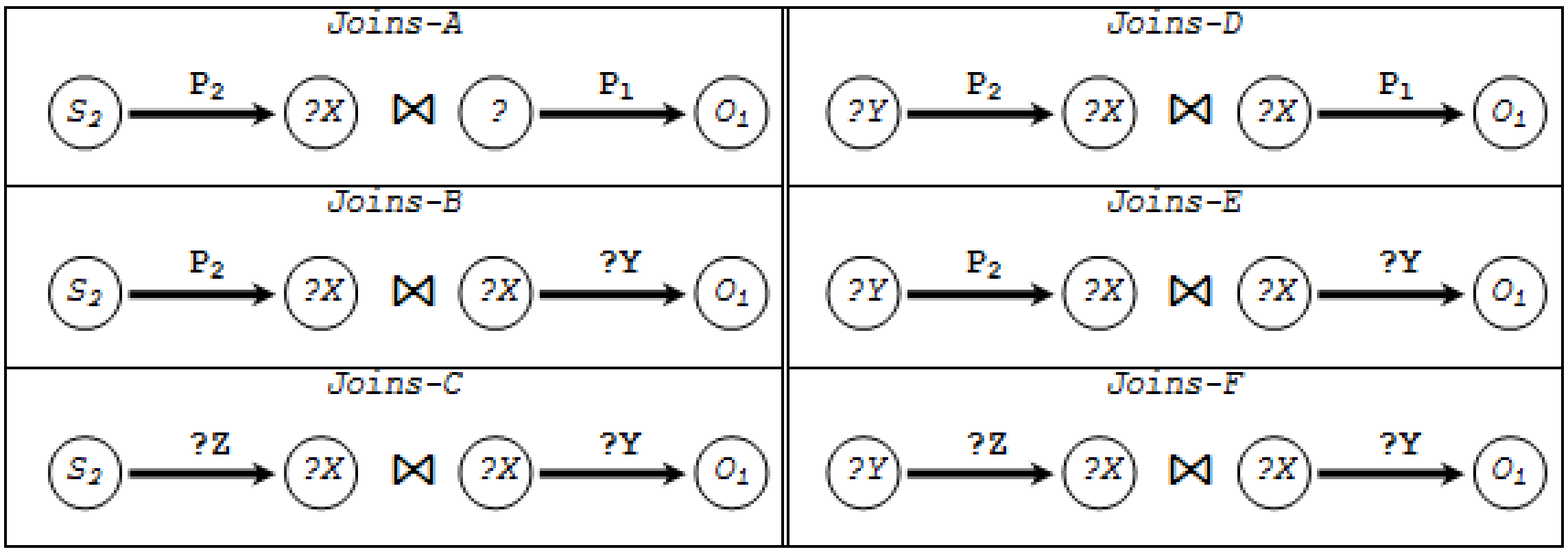

**Figure 4. Join categorization for Subject-Object (SO).**

$k^2$-triples currently supports joins subject-subject (SS) and object-objet (OO), and cross-joins between subjects and objects (SO). We classify these joins in six categories in accordance with their implementation on the $k^2$-tree (Figure 4 shows examples of each category for SO joins):

- Joins-A: is the simplest class and comprises queries in where only the join variables are unbounded. That is: (?X,$P_1$,$O_1$) (?X,$P_2$,$O_2$) for SS, ($S_1$,$P_1$,?X) ($S_2$,$P_2$,?X) for OO, and (?X,$P_1$,$O_1$) ($S_2$,$P_2$,?X) for SO. These are implemented by individually solving each triple pattern upon *direct* or *reverse neighbors* retrieval and the resulting lists are then intersected to obtain the final result. Note that this intersection is performed in a very faster way by taking advantage of the ID-ordered of both lists.

- `Joins-B`: are similar to the previous one but, in this case, one triple pattern have an unbounded predicate; (?X,?Y,$O_1$) ($S_2$,$P_2$,?X) is an example for `SO`. We firstly resolve the pattern with the bounded predicate and the resulting list is then intersected with all lists obtained from the pattern with unbounded predicate. The final result is the union of all these lists
- `Joins-C`: consists in joins in where both triple patterns have unbounded predicates: (?X,?Y,$O_1$) ($S_2$,?Z,?X). The process resolves each pattern and then intersects the resulting lists by taking advantage of their ID-ordering.

All these patterns have in common that predicates can be bounded or unbounded but non-joined subjects and objects (in each case) are given as constants. The next three categories consider non-joined unbounded subjects and objects:

- `Joins-D`: comprise patterns with bounded predicates and one subject or object variable in one pattern; (?X,$P_1$,$O_1$) (?Y,$P_2$,?X) is an example for `SO`. These joins resolve the less uncertainly pattern and retrieves all possible values for the join variable. In the previous example, a *reverse neighbors* retrieves all subjects matching with (?X,$P_1$,$O_1$). The second pattern is then transformed into a patterns group in where the join variable is bounded with all results retrieved in the previous step and each resulting pattern is individually resolve. This way, the pattern (?Y,$P_2$,?X) is resolved through patterns (?Y,$P_2$,$O_i$) in which $O_i$ are all results obtained for the initial pattern.
- `Joins-E`: are similar to the previous one but adding an unbounded predicate. They are solved by repeating the previous operation for each predicate in the dataset.
- `Joins-F`: add a second unbounded predicate to the conditions of `Joins-E` and are resolved by performing `|P|` times the operation for `Joins-E`.

## EVALUATION

We design our experimental setup using datasets from different domains[8] in order to obtain an heterogeneous corpora. All of them are previously cleaned to delete duplicated triples. Their resultant properties are described in Table 1: the column "Size" shows the original size in raw N3 format, "Triples" indicates the number of triples in the dataset and the three latest columns show the number of different subjects, predicates and objects respectively. We choose different-size datasets to analyze the impact of the number of triples in the results. Besides, we also consider the number of predicates to study the main bottleneck of vertical-partitioning.

We experiment against state-of-the-art solutions like RDF-3X, BitMat, MonetDB, and Hexastore, but the comparisons are only performed against the three first ones which allow larger datasets to be managed. We perform on MonetDB by following (Sidirourgos et al., 2008): we create separated predicate tables and insert triples with a subject-object order.

All tests are performed on ID-based representations for a fair comparison. We consider that the Dictionary is a common structure for these approaches, hence all of them could use any competitive technique to represent it. However, compact dictionaries are an open problem (Brisaboa, Cánovas, Martínez-Prieto and Navarro, 2011) out of the scope of this paper.

### Compression

This analysis focuses on the space required to represent the Triples component of an RDF dataset. Table 2 summarizes results for $k^2$-triples and the remaining techniques.

The comparison on the smallest dataset (*geonames*: 1GB on N3) already shows that the approaches on vertical-partitioning achieves the most compressed representations: $k^2$-triples and MonetDB take 0.017GB and 0.072GB. Both techniques largely outperform the multi-indexes based solutions which require 0.569GB and 0.344GB (for BitMat and RDF-3X respectively). These differences are kept for all datasets. Results for *dbpedia* are specially interesting by considering that it is a general-purpose dataset using a large set of predicates. We firstly emphasize that BitMat experiments on *dbpedia* did not finish. In turn, RDF-3X uses 9.529GB to index the dataset: this size is much greater than the used for MonetDB and $k^2$-triples (more than 8GB). Our approach takes ≈50% of the space used in MonetDB. These results demonstrate the $k^2$-triples ability to ultra-compress RDF, outperforming the multi-index based solutions in orders of GB.

---

[8] *geonames*, *dbtune* and *uniprot* are extracted from the BillionTriples Challenge 2010 (`http://km.aifb.kit.edu/projects/btc-2010/`), *wikipedia* is available at `http://labs.systemone.at/wikipedia3` and *dbpedia* at `http://wiki.dbpedia.org/Downloads36`.

| Dataset | Size (GB) | #Triples | #Subjects | #Predicates | #Objects |
|---|---|---|---|---|---|
| **geonames** | 1.00 | 9,415,253 | 2,203,561 | 20 | 3,031,664 |
| **wikipedia** | 6.72 | 47,054,407 | 2,162,189 | 9 | 8,268,864 |
| **dbtune** | 9.34 | 58,920,361 | 12,401,228 | 394 | 14,264,221 |
| **uniprot** | 9.11 | 72,460,981 | 12,188,927 | 126 | 9,084,674 |
| **dbpedia-en** | 33.12 | 232,542,405 | 18,425,128 | 39,672 | 65,200,769 |

**Table 1. Datasets description**

| Dataset | N3 Size | BitMat | RDF-3X | MonetDB | $k^2$-triples |
|---|---|---|---|---|---|
| **geonames** | 1.00 | 0.569 | 0.344 | 0.072 | 0.017 |
| **wikipedia** | 6.72 | 1.945 | 1.403 | 0.351 | 0.122 |
| **dbtune** | 9.34 | 6.072 | 2.033 | 0.443 | 0.149 |
| **uniprot** | 9.11 | 4.227 | 2.279 | 0.544 | 0.080 |
| **dbpedia-en** | 33.12 | × | 9.529 | 1.769 | 0.864 |

**Table 2. Compression results (in GB)**

### Querying

We design our experimental environment on a machine AMD-Phenom™-II X4 955@3.2GHz, quad-core (4cores-4siblings: 1thread per core), 8GB DDR2@800MHz. All tests are performed on the *dbpedia* dataset and the SPARQL queries are extracted from the log provided for the USEWOD'2011 Challenge[9]. This means that out testbed is composed by real user queries except for the patterns `(S,P,O)` which are directly extracted from the original dataset. All queries are translated to their correspondent IDs. Note that BitMat is not considered in these tests because of it does not finish its processing on *dbpedia*.

Table 3 shows averaged times (in milliseconds/pattern) for simple patterns. We discard `(?S,?P,?O)` by considering that this query is equivalent to a dataset dump. First, we compare against MonetDB to evaluate the performance of $k^2$-triples as a vertical-partitioning solution. The use of $k^2$-trees instead of tables largely improves query times for all patterns. This is specially significative for patterns with unbounded predicates because their resolution is the main weakness of the vertical-partitioning systems; for instance, `(S,?P,O)` takes 757 seconds/pattern in MonetDB and only 20.75 milliseconds/pattern in $k^2$-triples. These large differences are an experimental evidence of the $k^2$-tree high-performance for SPARQL querying.

The comparison of $k^2$-triples and RDF-3X also shows interesting conclusions. On the one hand, $k^2$-triples is the most efficient approach for all patterns with a given predicate. It means that the dual subject-object indexing of the $k^2$-tree outperforms the RDF-3X multi-indexes when the pattern binds the predicate. On the other hand, for patterns with unbounded predicates, RDF-3X is more efficient than $k^2$-triples. However the difference is largely reduced in comparison to MonetDB, even outperforming RDF-3X for `(S,?P,O)`. This fact proposes $k^2$-triples as a real choice even when the number of predicates grows to the order of thousands as occur in *dbpedia*. However, this is an extreme case as demonstrate the other datasets which use more limited-size predicate dictionaries. In these cases, $k^2$-triples is competitive with respect to multi-index approaches because of the cost of managing unbounded predicates is largely reduced.

Results in Table 4 complement previous experiments. We consider additional tests to measure the $k^2$-triples ability to perform join queries. We follow the six-category division reported in the previous section and consider 10 join queries in each one. All queries (also extracted from the USEWOD'2011 log) comprise two triple patterns to avoid the join performance to be influenced by the execution of query plans. The comparison of $k^2$-triples against MonetDB shows similar results than for triple patterns: our approach always overcomes the performance of the vertical-partitioning on relational tables. As we expected, times largely increase for all categories with any unbounded predicate. $k^2$-triples gets reasonable times for solving joins with a single unbounded predicate (`B` and `E`). However, these largely increase when the query comprises two unbounded predicates (`C` and `F`). Note that MonetDB results are not reported for these categories due to the very large times that it takes for their resolution. Comparisons with respect to RDF-3X join resolution demonstrate the superiority of $k^2$-triples for queries with no unbounded predicates. For the rest ones, RDF-3X is clearly superior by taking advantage of its multi-index.

[9] `http://data.semanticweb.org/usewod/2011/challenge.html`

| Pattern | (S,P,O) | (S,P,?O) | (S,?P,O) | (S,?P,?O) | (?S,P,O) | (?S,P,?O) | (?S,?P,O) |
|---|---|---|---|---|---|---|---|
| **RDF-3X** | 26.58 | 45.87 | 47.57 | **34.90** | 33.53 | 2668.46 | **34.55** |
| **MonetDB** | 26.14 | 50.29 | 757448.11 | 677111.48 | 97.49 | 6397.92 | 675127.13 |
| **$k^2$-triples** | **0.005** | **0.08** | **20.75** | 139.74 | **0.44** | **1297.49** | 49.92 |

**Table 3. Query times (in ms/pattern) for simple triple patterns**

| Pattern | Joins-A | Joins-B | Joins-C | Joins-D | Joins-E | Joins-F |
|---|---|---|---|---|---|---|
| **RDF-3X** | 122.43 | **81.78** | **58.99** | 68.92 | **4196.07** | **772.86** |
| **MonetDB** | 306.83 | $1691.85 \times 10^3$ | × | 60.26 | $1641.93 \times 10^3$ | × |
| **$k^2$-triples** | **97.09** | 126.47 | 3610.84 | **50.14** | 54295.54 | 13100.50 |

**Table 4. Query times (in ms/query) for join queries**

## DISCUSSION AND FUTURE WORK

This paper presents a novel compressed RDF engine ($k^2$-triples). It stores the RDF structure in an ultra-compressed representation which can be fully loaded and queried in main memory. This does not only result is space savings but also in competitive query times. The experimentation presented in this paper endorses these features by showing that: 1) $k^2$-triples is the most effective technique in between all considered solutions; 2) $k^2$-triples is the most efficient engine for solving triple patterns with bounded predicates and largely outperforms vertical-partitioning when the patterns include unbounded predicates. This achievement reinforces the vertical-partitioning opportunities as a competitive solution to perform on datasets with moderate amount of predicates; 3) the $k^2$-triples mechanisms for conjunctive queries take advantage of the triples pattern resolution to obtain the best performance for joins with no unbounded predicates.

All these experimental results propose $k^2$-triples as a competitive solution for full-in-memory RDF engines. To agree with its current performance, $k^2$-triples seems the best choice for representing datasets on specific knowledge areas which use a limited number of predicates. This fact motivates our future work which is drawn on several complementary issues. On the one hand, a query optimizer might allow more complex conjunctive queries to be efficiently resolved. On the other hand, we analyze $k^2$-triples to be enhanced to support indexed access on predicates. This decision leaves the vertical-partitioning philosophy but opens a new opportunity for ultra-compressed multi-indexes on $k^2$-trees. Additionally, the conversion of the current $k^2$-tree into a dynamic data structure would allow us to face up a more ambitious objective: the implementation of a fully-functional RDF-Store supporting not only querying but also insertion, updating and deletion of triples in the dataset.

## ACKNOWLEDGMENTS

This work is funded by the MICINN of Spain (PGE and FEDER) TIN2009-14560-C03-02, TIN2010-21246-C02-01 and CDTI CEN-20091048, Xunta Galicia (cofunded with FEDER) ref. 2010/17 (first and second authors), and by MICINN ref. BES-2010-039022 (FPI program), for thefirst author . MICINN TIN2009-14009-C02-02 (third and fourth authors), Junta de Castilla y León and the European Social Fund (third author) and Institute for Cell Dynamics and Biotechnology (ICDB), Grant ICM P05-001-F, Mideplan, Chile (fourth author).